# The observed age gradient in the Milky Way – as a test for theories of spiral arm structure

Jacques P Vallée

Herzberg Astronomy and Astrophysics Research Center, National Research Council of Canada
5071 West Saanich Road, Victoria, British Columbia, Canada V9E 2E7
ORCID  http://orcid.org/0000-0002-4833-4160

**Abstract.**
Some important predictions from 4 main models of spiral arm formation are tested here, using observational data acquired for the Milky Way galaxy.  Many spiral arm models (density wave, tidal wave, nuclear Lyapunov tube, or dynamic transient wave) have some consistencies with some of the observations, and some inconsistencies.

Our 4 tests consist of the relative **locations** and relative **speeds** of different arm tracers away from the dust lane, and the global arm **pitch angle** as obtained over two Galactic quadrants and several Galactic radii, as well as the arm's **continuity of shape** from Galactic quadrant IV to Galactic quadrant I.

In the Milky Way, an age gradient is observed from different arm tracers, amounting to **12.9 ±1.1 Myrs/kpc,** or a relative speed away from the dust lane of **76 ±10 km/s**.  The presence of an age gradient is predicted by the density waves, but is not consistent with the predictions of the tidal waves, of the nuclear Lyapunov tubes, nor of the dynamic transient recurrent waves.

1. **Introduction.**

Long periodic arms in spiral galaxies are well known, and they also exists in our Milky Way disk galaxy – for a review, see Vallée (2017b). Four main models compete for the formation of long spiral arms: density wave, tidal waves from a passing galaxy, Lyapunov tubes from the Galactic nucleus, dynamic transient waves, etc (see Khrapov et al 2021).

To define a spiral arm, some have looked at a *single* tracer (O-B stars, Open star Clusters, etc) at *different* ages to look for a color/age gradient.  Others have looked at *different* tracers, each tracer with a *different* well defined mean age (protostars, young  small radio HII regions with radio recombination lines, old optically-visible HII regions, etc). Open star clusters tend to be old: the age distribution of open star clusters shows none with a young age below 2 Myrs – see  Castro-Ginard et al (2021 – their Fig. 1); they no longer show a spiral arm beyond 30 Myrs (their Fig. 2), and are observed to reach an age beyond 10 Billion years (by which time they may wander in the next spiral arm).

What kind of specific arm tracers are needed ? Spiral arms are well observed through young stars (often bunched together). Excellent arm tracers are protostars and their masers (seen at radio wavelengths); other good tracers are HII regions (around young stars).

Where to look?  We need to look at the properties of the main long spiral arms, not the properties of the random stuff near the Sun. Many tracers can be found around the Sun, but the Sun is in a small Local Arm, also called an island armlet, located in the interarm between the main long Sagittarius and Perseus arms.

Young stars in the small Local Arm (an armlet, not a long log-periodic arm) could be born from multiple processes, and the Local Arm has different shapes in different tracers – see Vallée (2018a);  Laporte et al (2019). Rather than look at *nearby stars*, it might be useful to look at specific galactic regions (in galactic longitudes), such as where the line of sight from the Sun is *tangent to a long spiral arm*, to directly map the region from the inner arm side (with dust) across the spiral arms (to the outer arm side). Vallée (2014 – his Fig. 1 and Table 4) did that and found that each different arm tracer was located at a different distance from the inner arm edge, the younger tracers being closer to the inner arm edge (thus demonstrating gas flowing into the inner arm edge, and finding an age gradient). Also, alternating arms were grouped together, showing no discernable difference in their mean tracer offsets (his Fig.

3 and Table 5), thus removing the concept of a 'major' arm and a 'minor' arm. Vallée (2016a – his Fig. 2) grouped together the arms in Galactic Quadrant IV and showed that the mean properties of the arm tracers were the same (within the errors) as the mean properties of the arm tracers in the group of arms in Galactic Quadrant I. The arm tracer offsets in galactic longitude, going from Galactic Quadrant IV across the Galactic Meridian (Galactic longitude zero), became a 'mirror-image' of the arm tracer offsets in galactic longitude in Galactic Quadrant I. The observations (galactic longitudes of arm tangents) show the 4 main long log-shape arms are roughly equidistant in azimuth (along a circle around the Galactic Centre), thus not random in azimuth as for flocculent galaxies.

The author is aware of a rich wider context of studies of arm formation scenarios, both theoretical and observational, in nearby disk galaxies. My work on the Milky Way offers a better linear resolution (parsecs) and better sensitivity to some arm tracers (masers in starforming regions, say). A comparison with similar works in nearby spiral galaxies was made recentlyy, but with many arm tracers absent (due to a lack of sensitivity or linear resolution – see Table 1 in Vallée 2020a). Also, a recipe was added to select arm trcers and fitting functions to detect an offset between a starforming region near a 'shock front' and a non-starforming region near a 'potential minimum' of an arm (see Table 2 and Figure 11 in Vallée 2020a). As these two different regions of the same spiral arm are separated by about 350 pc, the linear resolution must be better than 70 pc (at the 5-sigma level).

1.1 Density waves, predicting an age gradient

Classic density-wave spiral arms were proposed by Lin & Shu (1964) and Lin et al (1969), predicting a dust lane (**D arm**) associated with young stars, and a 'potential minimum' (**P arm**) associated with old stars. They were slightly modified later to introduce a shock at the dust lane, by Roberts (1969) and Roberts (1975). Then Gitting & Clarke (2004 – their Fig.16) positioned the optically visible star-forming arm (OVSF arm, or **O arm** for short) for mid-age stars. In all these theories, gas moves across spiral arms (being faster than the spiral arm pattern, for gas lying below a galactic radius called co-rotation).

While the shock and the dust lane are very close together, and while the Potential Minimum and the old stars are close together, in between we have the protostars and radio masers (**M arm**), as very young stars or protostars may not become visible at optical wavelengths until much later.

We then expect an age sequence to be like: the D arm (dust and shock – age 0), the radio-detected maser star-forming arm (M arm, about age 0.8 Myrs), some radio-optical visible young stars with HII regions (O arm, 2 Myrs), the P arm (potential minimum, old stars), giving thus D-M-O-P.

Using radio wavelengths, Vallée (2014 – his Fig.1) noted a clear offset between the bright red dust arm (shock; D arm), the dust-hidden radio maser arm (M arm), then some very early compact HII regions, and finally the P arm where the broad diffuse CO 1-0 gas is found (with some old stars). Rather than taking specific physical tracers (dust, young stars, etc), some prefer to talk about waveband colours; at optical wavelengths, one can predict a red to blue change from the D arm to the O arm – see Yu & Ho (2018).

Tests on 24 nearby spiral galaxies (distance < 110 Mpc; redshift z < 0.028) indicated that the orbital angular speed of the gas (mean of 60 km/s/kpc) was higher than the arm pattern speed (mean of 36 km/s/kpc), consistent with the prediction of the density wave theory (Vallée 2020a). In those galaxies, one found a statistical separation around 370 pc between the P arm and the D arm. Recent theoretical reviews were done by Sellwood & Masters (2022) on self-excited instabilities for a recurrent cycle of groove modes, and Harsoula et al (2021) on precessing ellipses to build arms.

1.2 Tidal waves from nearby galaxies, and Lyapunov tubes from the Galactic Nucleus

Tides have been shown numerically to generate two spiral arms, but this effect is not long term. No age gradient is predicted. Dobbs & Baba (2014 – their Section 2.4) studied such tidal interactions and simulations (their Fig. 15).

Lyapunov tubes (Fig.4 in Romero-Gomez et al 2011) in the invariant Manifold theory (Fig. 4 in Romero-

Gomez et al 2006) have been proposed, producing arms and rings. A strong bar in the Galactic nucleus is employed. Some parts of the Manifold theory predicts a constant pattern speed, and are possibly generating shocks in the arms. Their arms are predicted to have various shapes and varying pitch angles with galactic radius (Fig. 2 in Athanassoula et al 2009), but they are not predicted to be shaped as 'log-periodic', and are not predicted to have a constant arm pitch angle with galactic radius. Recent tests of the predictions of the Manifold theory (between a strong nuclear bar and the arm pitch angle) by Lingard et al (2021 – their Section 3.2.2) would exclude the Manifold theory as a primary mechanism to drive the evolution of spiral arms.

### 1.3 Dynamic transient recurrent spiral arm model

The dynamic transient recurrent spiral model sprouted from the disaffectation from an earlier theory, as observational data (color gradient from the migration of OB stars away from the inner spiral arm shock) were hard to get to test the early density-wave theory (Shu 2016). This may be due to selecting optical wavelength bands (U, V, B, etc), each band being a composite of many different arm tracers. We need some specific arm tracers with a model of their evolution with time.

The 'dynamic transient recurrent' spiral arms are defined as transient and recurrent in nature (Dobbs & Baba 2014 – section 2.2). These arms can break up into smaller segments (of a few kpc) and reconnect (sheared by differential rotation) with other segments to reform large scale patterns (Pringle & Dobbs 2019). In this dynamic transient recurrent model, gas does not flow across spiral arms (the gas velocity being equal to the pattern speed), and the two (gas, pattern) are thus in co-rotation at each galactic radius (Baba et al 2016; Baba et al 2013). Thus global spiral arms may appear visually to be long-lived, yet are in fact assemblies of short-lived arm segments which break up often and then reconnect later with other arm segments, in an equilibrium between winding and self-gravity (Dobbs & Baba 2014).

Dynamic transient recurrent spiral arms exhibit co-rotation at every galactic radius (Dobbs & Baba 2014 – their Section 3.7). While there is no gas flow across a spiral arm, gas falls in from both sides of the spiral arm (e.g., Fig. 5 in Baba et al 2016). Thus gas falls into the spiral 'Potential Minimum' from both sides of a spiral arm - a kind of two large-scale colliding flows (Baba et al 2016). Gas can still undergo small shocks when falling into the potential minimum. For a small relative speed, there is no shock but a compression zone may emerge. For a larger relative speed, each arm could have two shocks or compressed zones, one at the inner arm edge, and one at the outer arm edge. This arm formation is said to be recurrent, giving some long-term arms.

Pettit et al (2015) showed a small change in pitch angle value with an increase in the galactic radius, namely that smaller pitch angle values are predicted nearer the Galactic Center (Fig. 10 b and 10c in Pettitt et al 2015). The dynamic transient recurrent spiral arm model predicts a large range of arm pitch angle, due to the winding of spiral arms as they turn around the Galactic nucleus, predicting a uniformity of the cotangent of the pitch angle values between two limits ($15°$ < pitch < $50°$, in Fig. 9 and Sect. 3.3 in Lingard et al 2021; Fig. 3 in Pringle & Dobbs, 2019).

Lingard et al (2021 – their Sect. 4) stressed that their observational data from 129 galaxies (redshift z < 0.055) are consistent with the dynamic transient arm model "if the minimum pitch angle is $15°$, but rule it out if the minimum pitch angle is $10°$." Lingard et al (2021) also noted that their statistical results about the cotangent of the pitch angle are "not evidence against the density wave theory, as the distribution of pitch angles may be dictated by other factors".

Elsewhere, Foyle et al (2011) used some tracers on 12 nearby galaxies and suggested that spiral arms are transitory.

The linear resolution and observational sensitivity in nearby galaxies (at megaparsec distances) is far worse than within the Milky Way (at kpc distances), affecting the exact location and detectability of a specific tracer (like a radio maser line in a starforming region, or like B-band emission).

In some galaxies, there could be a different pattern speed at different radial distances from the Galactic Center – see Rautiainen & Salo (1999) and Font et al (2014).

There are some variants to the density wave theory, notably the addition of swing amplification - see Dobbs & Baba (2014) and Font et al (2019).

Section 2 shows the spiral arm locations, as fitted earlier to arm tracers observed near the arm tangent.
Section 3 shows an age gradient, from using specific physical arm tracers (dust, etc), as a function of a linear offset away from the dust arm.
Section 4 shows the run of the global arm pitch angle, measured globally from data in two Galactic quadrants (IV and I), as a function of the Galactic radius.
Section 5 explores the map of the spiral arms beyond the Galactic Center. We conclude in summarising our tests, in Section 6.

2. **Arm locations from arm tangents – where are the dust lanes?**

Elsewhere (Fig. 1 in Vallée 2020b; Fig.2 in Vallée 2021a) we showed the 4 spiral arms in the Milky Way disc, fitted with a best pitch angle of -13.4$^o$, and each arm being equally spaced in azimuth by 90$^o$. The Sun is at 8.1 kpc from the Galactic Center. The match is done to the Galactic longitudes of the arm tangents as observed with the outer arm tracers, where the Potential Minimum is predicted.
These outer arm tracers include the diffuse CO arm tracer, the relativistic synchrotron electrons, the thermal electrons, and the middle aged HII regions – see Table 5 (diffuse CO), Table 6 (HII), Table 3 (thermal and synchrotron) in Vallée (2016a).

As discussed and published elsewhere, the start of each spiral arm near 2.2 kpc from the Galactic Center was found with the help of arm tangents as observed in Galactic quadrant IV and I – see Tables 4 and 5 in Vallée (2016b). The 3 known Galactic bars across the Galactic nucleus have been noted and discussed (Tables 2 and 3, Fig. 3 and Section 3.2 in Vallée 2016b), and the 'short boxy bar' (radius of 2.1 kpc) is retained there. Hence the 'long thin bar' was found to be unphysical, and this allows the spiral arm to go down to a Galactic radius near 2.2 kpc.

**Figure 1** shows our CO-fitted arm model in Galactic quadrant III. Also, we added the rotating gas loops predicted in the dynamic transient recurrent spiral arm model (Baba et al 2016 – their Fig.5), one on each side of a spiral arm, both impinging on the spiral arm. In their model, each gas loop is in motion (an epicyclic flow, with a guiding centre at a different galactic radius). Hence stellar spiral arms are formed by flows arriving from both sides of the spiral arm (their Section 3.3), bringing gas with them having opposite flow directions (their Figure 5). Thus this should create a single compressed zone in the middle of the arm (for very small relative flow speed), or else on each side of the arm (one at the inner arm side, and one at the outer arm side) for a higher relative flow speed (up to 15 km/s – see Section 3.2 in Wada et al, 2011). There should not be any shock (for a very small relative flow speed). In the dynamic transient spiral model, we would add one hot dust lane in the arm middle (relative flow speed <15 km/s), or two predicted hot dust lanes (higher relative flow speed), one on each side of each of the CO arm lane shown – see Figure 5 in Baba et al (2016). The 2 orange bars are not observed in the Milky Way, so this limits the gas flows to a maximum relative speed of 15 km/s.

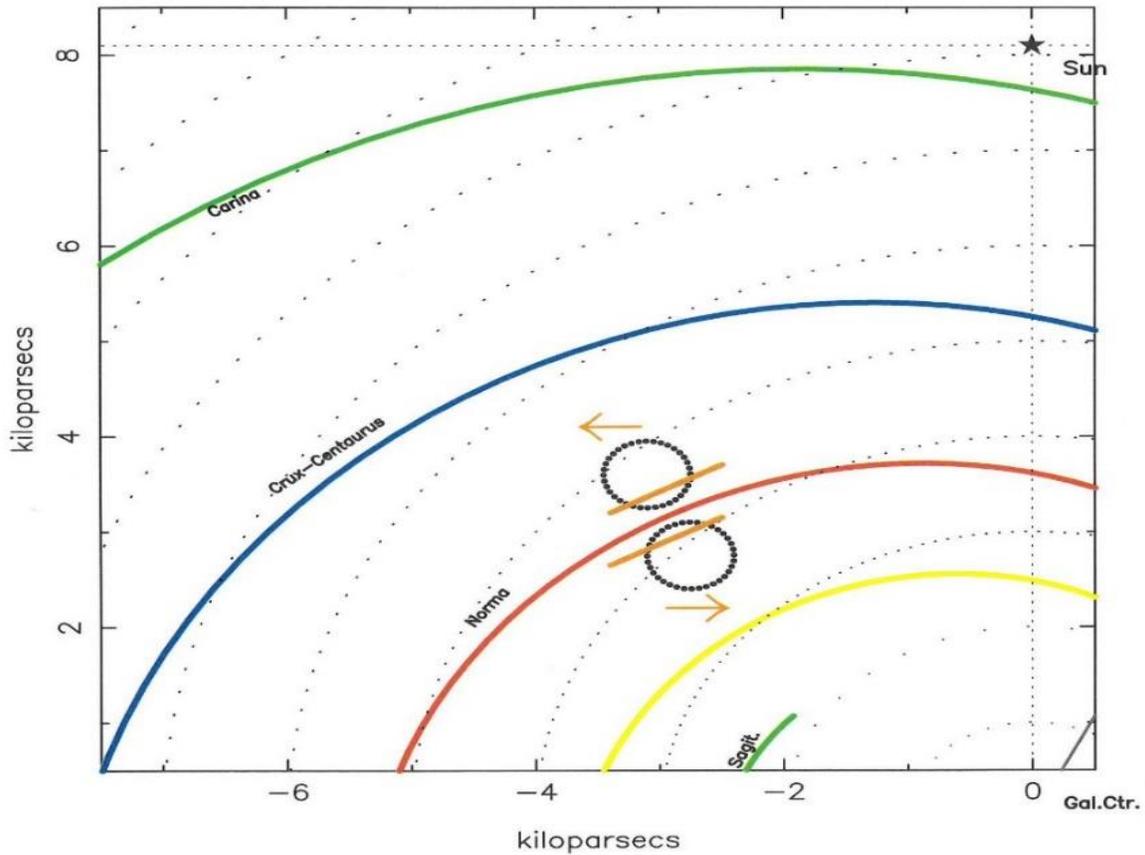

*Figure 1. Our CO-fitted model. In addition, as predicted in the dynamic transient spiral arm model (Fig. 5 in Baba et al 2016), each arm segment is produced in between two rotating loops, with the gas in each loop rotating counterclockwise (direction shown by an orange arrows). Thus the hot dust lane could be in the arm middle (red curve – using a very small relative flow speed up to 15 km/s following Wada et al 2011), or else could be on both arm sides (orange bars – using a more substantial relative flow speed).*

3. **An age gradient, from different arm tracers at different offsets from the hot dust lane**
   We built a figure of each arm tracer offset from the hot dust lane.

To build this figure, each observed arm tracer (CO, say) was collated from the literature into a catalog. For each arm tracer, in each spiral arm, a statistical mean longitude was made, along with a standard deviation of the mean longitude – see Table 5 in Vallée (2016a) for the diffuse CO peak longitude in each spiral arm segment, giving the statistical mean and s.d.m.; thus the mean and sdm for 10 observations of diffuse CO in the Norma arm is 328.4º ±0.8º. Ditto for the HII regions in each spiral arm, with its mean and s.d.m. (Table 6 in Vallée 2016a). Again this was done for other arm tracers, one table for each arm tracer. The conversion from angle to parsecs is explained also in Table 3 of Vallée (2016a): the distance from the Sun to the arm tangent at that Galactic Longitude is given there in kpc, and that distance is use to convert from angular to linear separation. An update set of tables for arm tracers is in preparation (Vallée, submitted).

In the density wave arm model (see Fig.2 in Roberts 1975), the shock with a gas density of 5 units (near the dust lane) is followed by a the potential minimum with a gas density near 2.5 units (a separation near 400 pc); after, the gas in orbit goes on to the potential maximum (in between two arms) with a gas density near 0.7 unit. In the Milky Way we observe that different arm tracers are separated in locations (Fig.1 in Vallée 2014; Fig.2 in Vallée 2016a). The observable dust lane is very close to the predicted location of the shock – Fig.4 in Roberts (1975) predicted a coincidence between the shock lane and the dust lane. The observable diffuse CO peak is very close to the predicted location of the potential minimum – Fig.4 in Roberts (1975) predicted diffuse HI (away from starforming sites) near the potential minimum, but here we prefer the nearby location of the diffuse CO peak (stronger peak, easier to observe than HI).

The mean age of the radio-observed radio masers in ultracompact HII regions has been calculated before, being near 0.7 Myrs (Xie et al 1996) and 0.4 Myrs (Section 4b in Wood & Churchwell 1989). For optically visible young stellar objects and compact HII regions, it is about 2.2 Myrs (Fig. 10 in Reggiani et al 2011) and 1.5 Myrs (Fig. 6e and Fig.7 in Hunt & Hirashita 2009). Given their different offsets from the dust lane or shock location, they provide an age gradient.

**Figure 2** shows an age gradient, using the physical separation of a tracer from the shock lane, and the age of that tracer at that particular separation (from a theoretical model). The red zone (including the shock lane, at left) is observed to be away by about 300 pc from the Potential Minimum (at center, x=0). Not all arm tracers are shown, for clarity (data in Vallée 2016a).

The **orange** zone with ultracompact HII regions and radio masers is pegged at roughly 1 ±0.6 Myr and about 150 pc from the Potential Minimum, or 150 pc from the dust/shock zone. Typical radio masers thus give a gradient or ratio of 1 Myrs/0.15 kpc = 7 ±4 Myrs/kpc;

The **green** zone with the radio recombination lines of young optically-visible HII regions is pegged at roughly 2 ±0.5 Myrs and about 100 pc from the Potential Minimum or 200 pc from the dust/shock lane. This gives an age gradient of 2 Myrs/0.20 kpc = 10 ±3 Myrs/kpc.

The **blue** zone with the diffuse broad CO gas and the old HII regions is pegged at roughly 4 Myrs and sitting at or very near the Potential Minimum, hence at about 300pc from the dust/shock zone. This gives an age gradient of 4 ±0.5 Myrs/0.30 kpc = 13 ±2 Myrs/kpc.

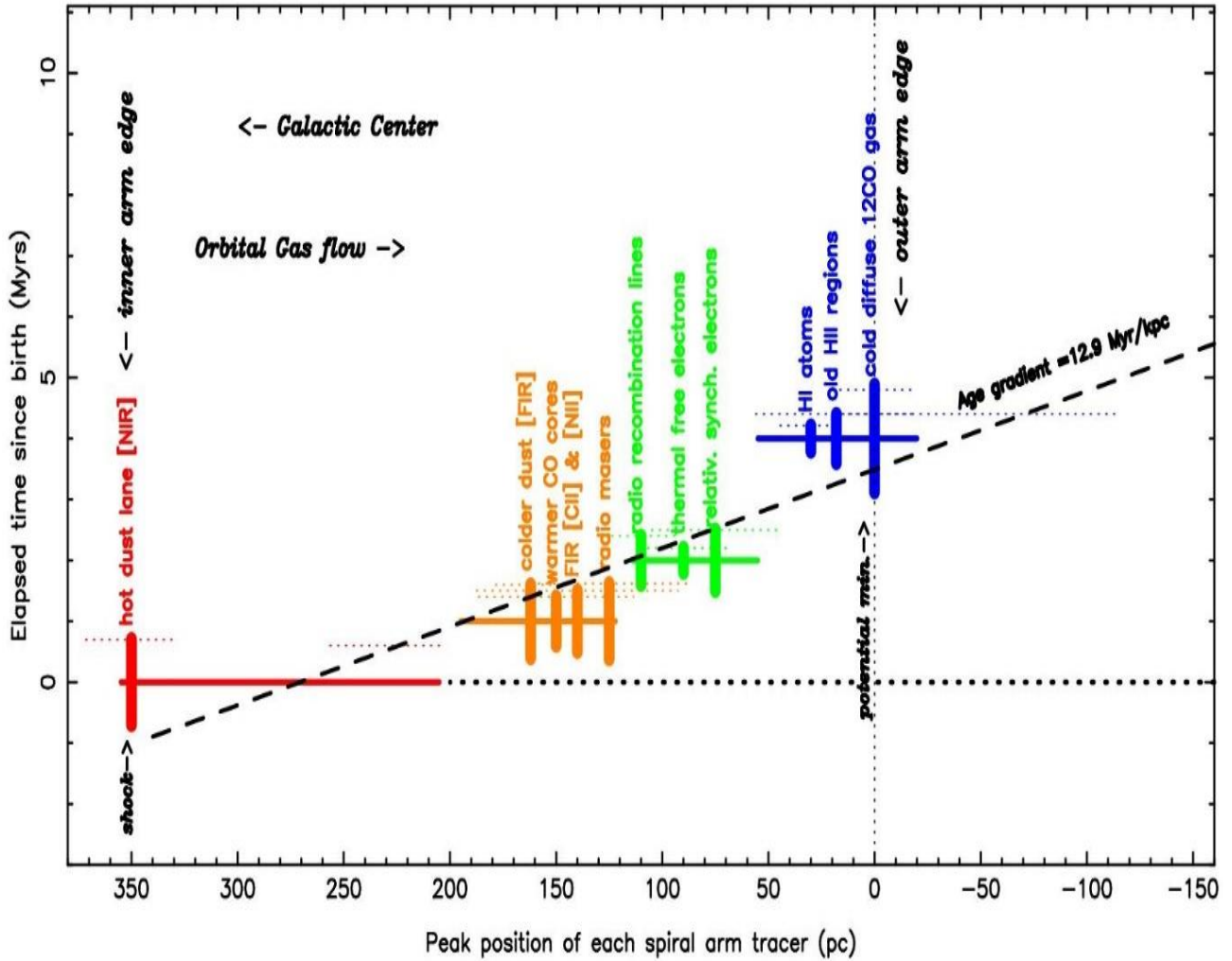

*Figure 2. Location of the observed separation of various arm tracers, as a function of distance from the shock lane (x-axis), and as a function of time elapsed since their birth (y-axis). The shock and dust (red **D-arm**), the radio masers (orange **M-arm**), and the potential minimum (blue **P-arm**) are explained in Section 1.1. The statistics were made to include 6 arm segments: Carina Crux-Centaurus, Norma, Perseus-start, Scutum, Sagittarius (see Table 1 in Vallée 2016a).*

 A *least-squares fitted line* was made through these points. The fitted observed age gradient shown (dashed line) is found by a least-squares fit to the arm tracers, yielding 12.9 ± 1.1 Myr/kpc, with a linear correlation coefficient of 0.99; the fit is done with 2 variables (time elapsed, and tracer separation) over 4 data (NIR/MIR origin/red, masers/orange, HII regions/green, CO/blue).
 The inverse of the age gradient gives the 'relative speed away from the dust/shock zone'. Hence a gradient of **12.9 ±1.1 Myr/kpc** can be inversed to give a relative speed of about 77.5 pc/Myr = **76 ±10 km/s.** These values confirm preliminary ones published in Vallée (2021b), using a means of a single variable.

 We can test for the offset between each arm tracer.

Test 1. In the density wave model, there is a sizeable offset gradient from the shock/dust lane to the Potential minimum (Lin et al 1969 – their Section 6). They predicted 3 'lanes', and we observed 4 coloured zones (Fig. 2 here) separated by about 350 pc. A typical CO tracer has a mean location error near 40 pc (Table 2 in Vallée 2017b), so the overall width of the 4 zones is a 8-sigma result (=350/40).

Test 1. In the model of tidal waves from passing galaxies, there is no prediction of an offset gradient of different arm tracers. No tracer offset gradient is predicted in model calculation either (Dobbs & Pringle 2010).

Test 1. In the Lyapunov tube model from a galactic nuclear bar, there is no prediction of an age gradient. No age gradient is predicted in bar-model model calculation either (Grant et al. 2012).

Test 1. In the dynamic transient recurrent spiral model (Baba et al 2016), there is no age gradient. The dynamic arm model predicts zero arm tracer offset. Some galaxies may show no offset, if the dynamic arm model operated there. For the arm separation seen in the Milky Way (Fig.1 in Vallée 2014; Fig.2 in Vallée 2016a), as summarized in Figure 2, the horizontal line at zero age (black dots) is the prediction of the dynamic spiral model (0 Myr/kpc).

We can test for the relative speed of an arm tracer, away from the dust/shock lane.

Test 2. In the density wave model, there is a sizeable relative velocity (Lin et al 1969 – their Section 6), necessary to create a shock lane. The global observational data for the Milky Way (Figure 2) indicate an age gradient, corresponding to a relative speed away from the shock/dust lane of 76 ±10 km/s (thus a 7-sigma result).

Test 2. In the model of tidal waves from passing galaxies, there is no prediction of an age gradient. No age gradient is predicted in model calculation either (Dobbs & Pringle 2010).

Test 2. In the Lyapunov tube model from a galactic bar, there is no prediction of an age gradient. No age gradient is predicted in bar-model model calculation either (Grant et al. 2012). No relative speed is predicted among arm tracers.

Test 2. In the dynamic transient recurrent spiral model (Baba et al 2016), there is no age gradient. There is a maximum relative flow speed up to 15 km/s in the theory of Wada et al (2011), creating a compression zone (not a shock). Our observed relative speed (76 ±10 km/s) *exceeds by a factor of 5* (=76/15) the maximum relative compression speed of 15 km/s predicted in the dynamic transient model of Wada et al (2011).

4. **Arm pitch angle along the galactic radius**

In the context of comparing the Milky Way to the global properties of nearby spiral galaxies, to advance our understanding of star formation and other themes, global parameters of the Milky Way are needed. Thus, for other nearby spiral galaxies, global values were listed elsewhere (Table 1 and Figure 1 in Davis et al 2017), and then compared, for correlating properties such as their mean arm pitch angle versus their central supermassive black hole.

While parts of a spiral arm may give a local pitch angle, here we want a global pitch angle value as fitted over a whole arm and observed in both Galactic quadrants IV and I together. The global arm pitch angle fits the whole arm, located in two Galactic quadrants (IV and I). This pitch angle is fitted separately for several arm tracers (CO, HII, masers, etc), and a mean pitch is obtained for all tracers - see Table 1 in Vallée (2017a) for Norma, and see Vallée (2015) for the Carina-Sagittarius arm (his Table 1) and for the Crux-Scutum arm (his Table 2).

One should stay away from a method that covers a very small area of a spiral arm, as their pitch angle results seem to oscillate - see table I in Vallée (2017c) for the observed pitch angles of each spiral arm, where the pitch angle for short sections of the Sagittarius-Carina arm is given from -7° to -15°. For a global pitch angle, the method to measure the pitch angle should cover a huge area of a spiral arm. See figure 1 covering two Galactic Quadrants, in Vallée (2015), and the results in Table 1 there giving -13.4° for the star-forming tracers in the Sagittarius-Carina arm, using Equations 1 to 10 there.

**Figure 3** shows, for each spiral arm, the location of the arm tangents from the model (yellow), the broad CO (blue; see Table 3 in Vallée 2016a), and the hot dust (red; see Table 2 in Vallée 2017b). In essence, this figure shows the fit of the model arms (predicted arm tangents) to the observations of the arm tangents (in Galactic longitude) for the diffuse CO 1-0 (Table 3 in Vallée 2016a) and the hot dust (Table 2 in Vallée 2017b). The x-coordinate and y-coordinates are both expressed in degrees – an excellent fit would yield the observed CO tracer to be on top of the model yellow tracer, hence 0° of separation.

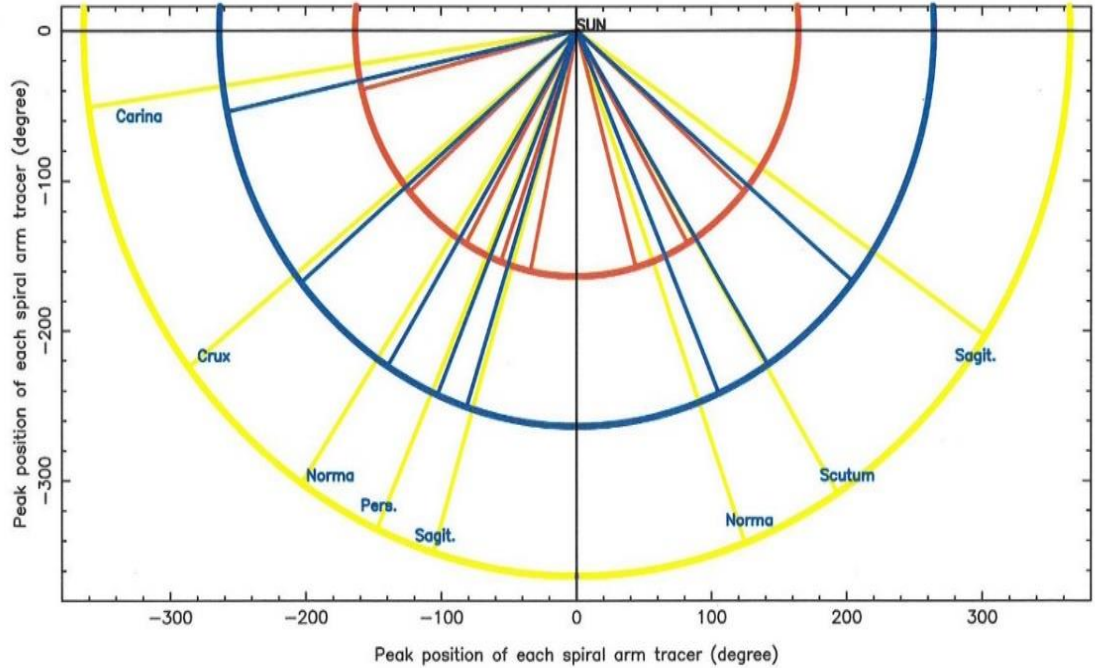

*Figure 3. Locations of arm tangents: (yellow) computed model, (blue) observed broad diffuse CO gas, and (red) observed hot dust lane, for each long spiral arm. It can be seen that the hot dust arm tangent is, in each arm, always closer to the Galactic Meridian (vertical line at 0° of Galactic longitude) than the broad diffuse CO arm tangent in that arm. The blue tangent is essentially where the 'Potential Minimum' is predicted, while the red tangent is essentially where the compressed or shocked lane is predicted.* From our CO-fitted model, our computed CO arm tangents were read at Galactic longitudes -83° (Carina), -52° (Crux), -34° (Norma), -24° (Perseus start), -17° (Sagittarius start), +20° (Norma), +32° (Scutum), and +57° (Sagittarius).

The quality of the fit, the difference between our best CO-fitted model and the observed arm tangents, is about -0.2° ±1.6° - see **Figure 3** here. Comparing Galactic longitudes between the model tangents and the arm tangents to the outer arm tracers, one would expect a zero offset, and indeed one finds an offset that is not significant – below the 0.1-sigma level (=0.2/1.6). Such an angular error of 0.2° corresponds to a linear error of 21 pc, at a typical distance of about 6 kpc from the Sun. Also, the fact that each tracer separation from the dust lane in Galactic quadrant IV (Fig. 2a in Vallée 2016a) is practically the same in Galactic quadrant I (Fig. 2b in Vallée 2016a) is proof of a very small statistical error, otherwise the two figures (quadrants) could not match.

The fit of the am model is done to the broad diffuse CO tangents and outer arm tracers. The separation of the yellow model longitude (outer arm tracers) from the red hot dust longitude (away from the Galactic Meridian) has an average of 4.7° with a s.d.m. of 0.7° (significant above the 6-sigma level). At a mean distance to the Galactic Center of 5.0 kpc, this angular offset amounts to 410 pc.

**Figure 4** shows the observed run of the pitch angle, separately for the arm tracers near the dust lane (inner arm tracers - Fig. 4a), and for the arm tracers near the broad CO lane (outer arm tracers - Fig.4b).

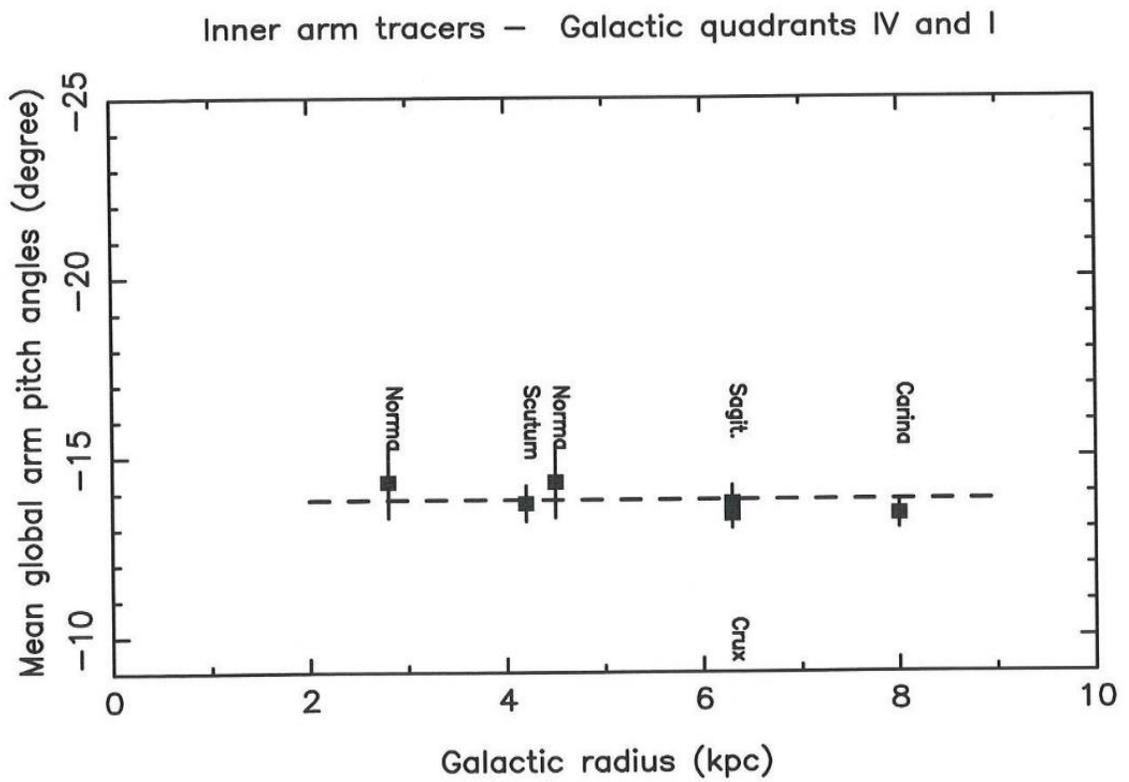

*Fig. 4a. Using only the inner arm tracers, this figure shows the run of the observed global arm pitch angle as a function of the Galactic radius (black squares). The mean observed run is seen in the black horizontal dashed line at pitch = -13.8º. Inner arm tracers used included: dust lane (240 µm), maser lane, FIR [CII] and [NII] gas.*

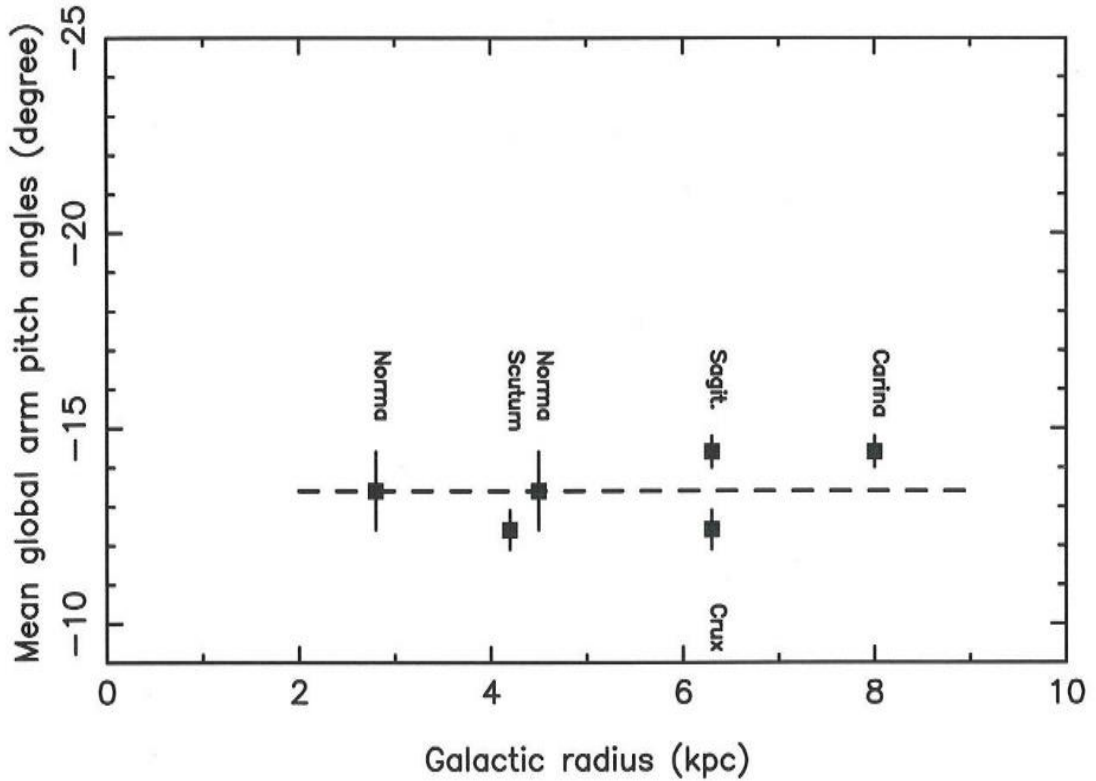

*Fig. 4b. Using only the outer arm tracers, this figure shows the run of the observed global arm pitch angle as a function of the Galactic radius (black squares). The mean observed run seen is the black horizontal dashed line at pitch angle = -13.4°. Outer arm tracers used included: middle aged HII regions, relativistic synchrotron electrons, thermal electrons, broad diffuse CO gas.*

One can see no obvious shift in pitch angle between the inner arm pitch (from new stars and dust) at -13.8° ±0.2° and the outer arm pitch (from old stars and broad diffuse CO) at -13.4° ±0.6°, giving a difference of only 0.4° ±0.6° (not statistically significant).

Summarizing for the Milky Way, the pitch angle values, measured with data from two Galactic Quadrants, seem to be constant as one goes from the Galactic Center to the Sun's orbit (no change of pitch angle with increasing galactic radius – see Fig. 4 here). Using a model of the Milky Way arms with a unique pitch angle for all logarithmic spiral arms, one can extrapolate that model beyond the Galactic Center and up in Galactic Quadrants II and III; doing that can allow the comparison of predicted and observed locations (in kpc) of giant HII regions in the Perseus arm. This was done, and they agreed well – see Fig.1 in Vallée (2019).

Also, in the nearby disk galaxy M51, it was observed that the arm pitch angle is almost the same at all galactic radii from 1 to 10 kpc (Fig. 1 in Vallée 2016b): the running mean pitch appears flat with increasing galactic radius, with localized deviations never exceeding 20°.

We can test for the global arm pitch angle, as a function of the Galactic radius.

Test 3. In the density wave model, the pitch angle predicted matches the pitch angle observed (flat horizontal line), as the pitch angle is independent of Galactic radius (e.g., Pringle & Dobbs 2019). In the Milky Way and in M51, a flat line is observed.

Test 3. In the model of tidal waves from passing galaxies, there is no prediction of a constant pitch angle with galactic radius. Tidal arms are deformed, away from their Galactic Nucleus.

Test 3. In the Lyapunov tube model, there is no prediction of a constant pitch angle with galactic radius. Arms are deformed.

Test 3. The dynamic transient recurrent arm theory allows to have a diverse range of pitch angle at a single snapshot; it may also depend on the shear of the rotation curve. Thus the pitch angle predicted could change a little with Galactic radius, due also in part to the winding down of the arms nearer the Galactic Nucleus.

5. **Map of the outer side of the Galaxy (beyond the Galactic Center)**

In the process of doing this paper, we produced a map of the far side of the Milky Way disk (beyond the Galactic Center) similarly to what is done in Figure 1. It was fitted to the best pitch angle of the outer tracers in the observed spiral arms (diffuse CO gas, middle-aged HII regions, relativistic synchrotron and thermal electrons in two Galactic quadrants (IV and I). We fitted the best fitted arm pitch angle over the whole map (ignoring local perturbations and interarm islands). Since these arms are fitted in both Galactic quadrants *between* the Sun and the Galactic Center, their predicted locations *beyond* the Galactic Center should be quite accurate. Our arm model also fits the exact galactic longitudes of the well observed arm tangents over their long line of sight (good to $0.5°$, or 44 pc at a galactic radius of 5 kpc), providing a very good accuracy.

The part beyond the Galactic nucleus is extrapolated, and it should be very good, since its debut in Galactic Quadrants I-II (Norma arm, Scutum arm, Sagittarius arm) had to match the observational data in Galactic quadrants III-IV (Norma arm, Carina arm, Crux-Centaurus).

**Figure 5** shows this map, extending below the Galactic Center by about 20 kpc. Our arm pitch angle is global (not changed along the log-periodic arm). As above, the fit is done to the tangents (Galactic longitudes) of the diffuse CO arm tracer, the synchrotron electrons, the thermal electrons, and the middle aged HII regions – see Table 5 (diffuse CO), Table 6 (HII), Table 3 (thermal and synchrotron) in Vallée (2016a). The difference between the best model and the observed tangents), is about $-0.2° \pm 1.6°$ - see Figure 3 for the quality of the fit.

The shape of the arms taken is a log-spiral, a simple mathematical function as employed already by many authors to fit the observations of spiral arms in other nearby spiral galaxies – see an observational review in Vallée (2017b). We assume that the Milky Way galaxy is similar to most other nearby spiral galaxies. Island armlets and filaments between the arms exist, but are not fitted here. Non log-spiral shapes or a segmented arm shape could be accommodated later, for the minority of observed galaxies with such an appearance. An odd shape would add to the complexity of the mathematical analysis, and its final conclusion. There is not enough distant data from observations within our Galaxy to justify segmented arms (Occam's razor).

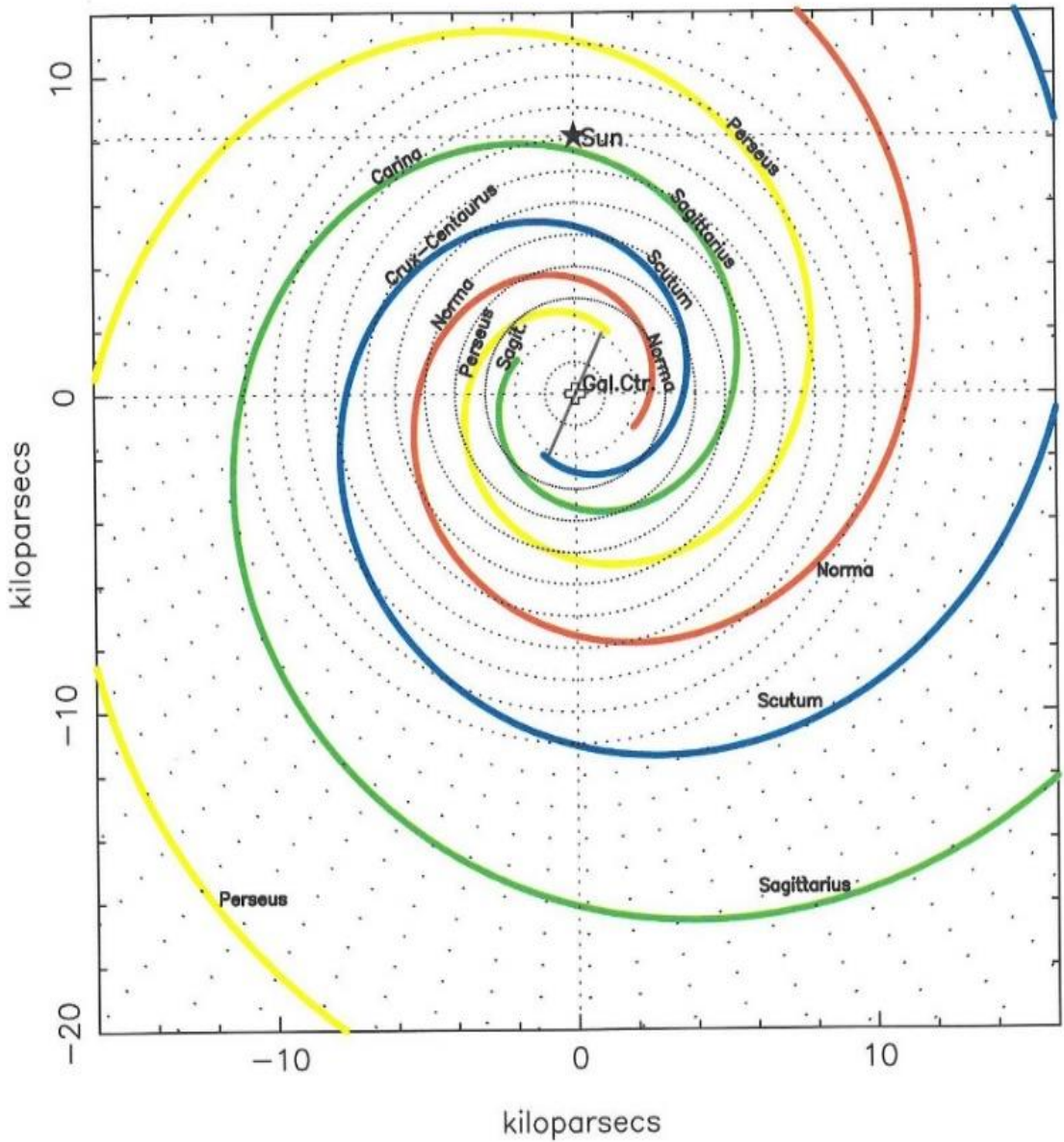

*Figure 5. Same as Figure 1, but our CO-fitted map extending way below the Galactic Center (Galactic quadrants I and IV).*

In contrast, others have predicted the map of the arms beyond the Galactic Center. Thus the map in Fig. 5 in Minniti et al (2021) employed the locations of classical Cepheids (optical and near-infrared photometric distances; typical errors near 1.0 kpc, rising to 2.2 kpc) while the arm widths near the Sun were measured to be near 0.4 kpc (see Fig. 2 in Vallée 2020b). The classical Cepheids can originate in small 'interarm islands' (not just in long arms – Vallée 2020c) confuse the issue of where to draw the spiral arms; an interarm Cepheid should be ignored when using Cepheids to map the long arms. They also made local adjustments to their model arms (changing the arm pitch angle to connect with the assumed nuclear bar location or another nearby arm location, not remaining log-periodic).

Similarly, the map in Fig. 2 in Xu et al (2021) employed the parallactic distances of masers (at radio wavelengths; typical total errors are near 2 kpc, rising to 3 kpc). The masers can originate in interarm islands (not just in the log-periodic long spiral arms) confuse the issue of drawing the arm locations; an interarm maser should be ignored when using masers to map the long arms. Their Fig.1 and 2 also proposed a bottom-up model, assembling numerous different arm segments, deviating from the log-normal seen in other spiral galaxies (their arm pitch angles were changed at various places, not remaining log-periodic).

We can text for the continuity of the arm shape, over several Galactic quadrants.

Test 4. In the density wave, the spiral arms continue beyond the Galactic Center to follow a log-spiral, as above (Fig.5). As mentioned earlier, the computed part beyond the Galactic Center has to match the observational data (it does match them).

Test 4. In the model of tidal waves from passing galaxies, there is no prediction of a constant log-spiral over galactic-wide scales and beyond the Galactic Nucleus.

Test 4. In the Lyapunov tube model, there is no prediction of a constant log-spiral over galactic-wide scales.

Test 4. The dynamic transient model follows semi-irregular recurring arms (colliding flows, secondary spurs), not regular log-spiral arms over several kpc. Thus Fig. 2 in Baba et al (2016) shows a different number of arms above and below the Galactic Center at some times (2.716 and 2.776 Gyrs).

6. **Conclusion.**

We tested 4 models for the formation of spiral arms, employing 4 observational data from the Milky Way (separation of arm tracers from each other at the arm tangents; relative speed of tracers from the dust lane; arm pitch angle at various galactic radius; continuation of arm shape over galactic scales).

Relative Arm tracer locations. Observations show a growing offset, away from the inner arm edge, of different arm tracers, and an age gradient can be observed among different arm tracers (Fig. 2). The observations of arm tracers show an age gradient of 12.9 ±2 Myr/kpc in the Milky Way, None is predicted in models, except in the density-wave model.

Relative arm tracer speed. The observed age gradient corresponds to a relative velocity of tracers away from the shock front of 76 ±10 km/s. The density-wave theory predicts a high enough speed to create a shock, while the dynamic transient recurrent model admits a very low relative tracer speed - enough to create a compression zone, not a shock (Fig. 1).

Pitch angle. Observations show that the global pitch angle of the spiral arm is the same at every galactic radius (Fig.4). No increase in pitch angle is predicted with galactic radius in the density wave theory, but some increase or decrease is allowed in all the other spiral arm models.

Arm shape. The constancy of the arm shape over galactic scales is predicted in the density wave theory. With a single global pitch angle, our model (Fig. 5) is fitted to an arm tracer in Galactic Quadrant IV, and the logarithmic spiral thus fitted is extended beyond the Galactic Center and comes back up to fit the arm tracer in Galactic Quadrant I. However, several changes in arm shapes and pitch angle are allowed in all the other spiral arm models.

Given the above, the density wave theory seems to prevail over all 4 tests and over some other theories, when testing with the observations of the Milky Way disk (good linear resolution).

What else the density wave theory can predict ? The 'angular spiral pattern speed' $\Omega_p$ and the concomittent corotation radius $r_{coro}$ have to be determined for each spiral galaxy. For the Milky Way galaxy, a table of the 'angular spiral pattern speed' $\Omega_p$ shows low values between 16 to 23 km/s/kpc (Table 2 in Vallée 2018b), and high values between 24 and 30 km/s/kpc (Section 1 in Vallée 2021a). These higher values for $\Omega_p$ mostly employ nearby optical stars from Gaia, but most of these local stars are not located in a long log-spiral arm caused by a density wave (they are in an island armlet, created differently, and thus should not be employed to get the density wave parameters). These lower values for $\Omega_p$, using arm tracers (radio masers, young optical HII regions) as separated from the dust lane, put the angular spiral pattern speed $\Omega_p$ between 12 and 17 km/s/kpc (Table 1 and Equation 1 in Vallée 2021a). A list of recent other measurements giving similarly low values of $\Omega_p$ is given elsewhere (Section 3 in Vallée 2021a), ranging from 12 to 20 km/s/kpc, giving a mean angular spiral pattern speed $\Omega_p$ of 16 ±4 km/s/kpc; since the observed stellar and gas circular rotation speed is near 233 km/s, it follows a corotation radius $r_{coro}$ of 15 ±4 kpc.

The increasing reach of precise observational data offers better tests of theoretical models. More accurate samples of the Milky Way over increasing distances may offer more stringent tests of arm formation theories - more recent complex variants of theoretical models could soon be tested.

**Acknowledgements.**

Figure production made use of the PGPLOT software at the NRC Herzberg Astro & Astro Research Centre in Victoria, BC. I thank a referee for helpful comments.**References**

15ranging from 12 to 20 km/s/kpc, giving a mean angular spiral pattern speed $\Omega_p$ of 16 ±4 km/s/kpc; since the observed stellar and gas circular rotation speed is near 233 km/s, it follows a corotation radius $r_{coro}$ of 15 ±4 kpc.

The increasing reach of precise observational data offers better tests of theoretical models. More accurate samples of the Milky Way over increasing distances may offer more stringent tests of arm formation theories - more recent complex variants of theoretical models could soon be tested.

**Acknowledgements.**

Figure production made use of the PGPLOT software at the NRC Herzberg Astro & Astro Research Centre in Victoria, BC. I thank a referee for helpful comments.

**References**

Athanassoula, E., Romero-Gomez, M., Bosma, A., Masdemont, J. 2009. Rings and spirals in barred
    galaxies. II. Ring and spiral morphology. MNRAS, 400, 1706-1720.
Baba, J., Saitoh, T., Wada, K. 2013. Dynamics of non-steady spiral arms I disk galaxies. ApJ, 763, 46 (1-1 4).
Baba, J., Morokuma-Matsui, K., Miyamoto, Y., et al 2016. Gas velocity patterns in simulated galaxies :
    observational diagnostics of spiral structure theories. MNRAS, 460, 2472-2481.
Castro-Ginard, A., McMillan, P., Luri, X., Jordi, C. et al. 2021. On the Milky Way spiral arms
    from open clusters in Gaia EDR3. A&A, 652, 162 (1-9).
Davis, B.L., Graham, A.W. Seigar, M.S. 2017. Updating the (supermassive black hole mass) – (spiral arm pitch angle) relation: a
    strong correlation for galaxies with pseudobulges. MNRAS, 471, 2187-2203.
Dobbs, C., Baba, J. 2014. Dawes Review 4: Spiral structures in disc galaxies. Publ. Astr. Soc. Australia, 31, e035 (1-40).
Dobbs, C.L., Pringle, J.E. 2010. Age distributions of star clusters in spiral and barred galaxies as a test for theories of spiral
    structure. MNRAS, 409, 396-404.
Font, J.,Beckman, J.E., Querejeta, M., et al 2014, Interlocking resonance patterns in Galaxy
    disks. ApJ Supp 210, 2, (1-30).
Font, J., Beckman, J.E., James, P., et al. 2019 Spiral structures in barred galaxies.
    Observational constraints to spiral arm formation mechanisms. MNRAS, 482, 5362-
    5378.
Foyle, K., Rix, H.-W., Dobbs, C., et al. 2011, Observational evidence against long-lived
    spiral arms in galaxies. ApJ 735, 101 (1-11).
Gittins,D.M., Clarke, C.J. 2004. Constraining corotation from shocks in tightly wound spiral galaxies. MNRAS, 349, 909-921.
Grand, R., Kawata, D., Cropper, M. 2012. Dynamics of stars around spiral arms in an N-body/SPH simulated barred spiral
    galaxy. MNRAS, 426, 167-180.
Harsoula,M., Zouloumi, K., et al. 2021. Precessing ellipses as the building blocks of spiral arms. A&A,
    655, a55, p1-17.
Hunt, I.K., Hirashita, H. 2009. The size-density relation of extragalactic HII regions. A&A, 507, 1327-
    1343.
Khrapov, S., Khoperskov, A., Korchagin, V. 2021. Modelling of spiral structure in a multi-component Milky Way-like galaxy.
    Galaxies, 9, 29 (1-28).
Laporte, C., Minchev, I., Johnston, K., et al 2019. Footprints of the Sagittarius Dwarf galaxy in the Gaia data set. MNRAS, 485,
    3134-3152.
Lin, C.C., Shu. F.H. 1964. On the spiral structure of disk galaxies. ApJ, 140, 646-655.
Lin, C.C., Yuan, C., Shu, F.H. 1969, On the spiral structures of disk galaxies. III. Comparisons with observations. ApJ, 155,
    721-746.
Lingard, T., Masters, K., Krawczyk, Lintott, C., et al . 2021, Galaxy zoo builder: Morphological dependence of spiral galaxy with
    pitch angle. MNRAS, 504, 3364-3374.
Minniti, J.H., Zoccali, M., Rojas-Arrigada, A., et al. 2021. Using classical Cepheids to study the far side of the Milky Way disk.
    Astron. & Astrophys., in press (arxiv.org/abs/2107.03464).
Pettitt, A., Dobbs, C., Acreman, D., Bate, M. 2015, The morphology of the Milky Way – II. Reconstructing
    CO maps from disk galaxies with live stellar distribution. MNRAS, 449, 3911-3926.
Pringle, J.E., Dobbs, C.L. 2019, The evolution of pitch angles of spiral arms. MNRAS, 490, 1470-1473.
Rautiainen, P., Salo, H. 1999, Multiple pattern speeds in barred galaxies. I. Two dimensional models. A & A, 348, 737-754.
Reggiani, M., Robberto, M., Da Rio, N., Meyer, M.R., et al 2011. Quantitative evidence of an intrinsic luminosity spread in the
    Orion nebula cluster. A&A, 534, A83 (1-12).